\documentclass{PoS}
\usepackage{graphicx}

\title{Discovery of VHE and HE emission from the blazar 1ES~0414+009 with H.E.S.S and Fermi-LAT}

\ShortTitle{Discovery of VHE and HE emission from the blazar 1ES~0414+009 with H.E.S.S and Fermi-LAT}

\author{F. Volpe$^a$, S. Ohm$^a$, M. Hauser$^b$, S. Kaufmann$^b$,  L. G\'erard $^c$ on behalf of the HESS collaboration\\
\llap{$^a$} Max-Planck-Institut f\"ur Kernphysik, Heidelberg, Germany\\
\llap{$^b$} Landessternwarte, Universit\"at Heidelberg, K\"onigstuhl, Heidelberg, Germany\\
\llap{$^c$} Astroparticule et Cosmologie (APC), CNRS, Universite Paris 7 Denis Diderot, Paris, France\\
E-mail: \email{francesca.volpe@mpi-hd.mpg.de}}

\author{L.~Costamante$^d$, S.~Fegan$^e$, M.~Ajello$^d$ on behalf of the Fermi-LAT
Collaboration \\
\llap{$^d$} W. W. Hansen Experimental Physics Laboratory, Kavli Institute for Particle Astrophysics and Cosmology, Department of Physics and SLAC National Accelerator Laboratory, Stanford University, Stanford, USA\\
\llap{$^e$} Laboratoire Leprince-Ringuet, Ecole Polytechnique, CNRS/IN2P3, Palaiseau\\
}

\abstract{The high energy peaked BL Lac (HBL) object 1ES~0414$+$009 (z=0.287) is a distant very high-energy (VHE, E~$>$~100 GeV) blazars with well-determined redshift. This source was detected with the High Energy Stereoscopic System (H.E.S.S.) between October 2005 and September 2009. It was also detected with the Fermi Large Area Telescope (LAT) in 21 months of data. The combined high energy (HE) and VHE spectra, once corrected for $\gamma$-$\gamma$ absorption on the extragalactic background light (EBL), indicate a Compton peak located above few TeV, among the highest in the BL Lac class.}

\FullConference{25th Texas Symposium on Relativistic Astrophysics - TEXAS 2010\\
		December 06-10, 2010\\
		Heidelberg, Germany}

\begin{document}

\section{Introduction} \label{intro}

The BL Lac object 1ES~0414+009 is located a redshift of z~=~0.287 \cite{halpern91} and 
harbors a super-massive black hole of mass $\sim 2 \times 10^9$ M$_\odot$ \cite{urry2000}.  
The host galaxy is classified as elliptical, with luminosity -23.5 \cite{falomo93}.
This source belongs to the class of high-frequency-peaked BL Lacs (HBL), 
objects with a synchrotron-emission peak located at UV/soft-X-ray
frequencies.

Based on the VHE estimates taken from \cite{costa_ghisellini}, H.E.S.S. started 
to observe 1ES~0414+009 in 2005 and continued until 2009, 
dedicating significant observing time to 
its detection, because its high redshift made this source an interesting candidate for EBL studies.
Results of the H.E.S.S. measurements will be here reported, together with observations carried out in these
5 years by other
instruments in different wavelenghts.
Indeed, the source is also detected in the HE (100 MeV $-$ 100 GeV) range
in the first 20 months of operation of the Large Area Telescope (LAT) 
on the \textit{Fermi} $\gamma$-ray space telescope. 
X-ray and UV observations carried out by the \textit{Swift} XRT and UVOT 
instruments and optical measurement from the ATOM telescope complete 
the broad-band spectral energy distribution (SED) 
of 1ES~0414+009. Properties of the SED and constraints on the EBL are also discussed in the following.

\section{H.E.S.S. observations}\label{HESSDATA}

H.E.S.S. has observed the HBL 1ES~0414+009 every year between 2005 and 2009, 
accumulating 73.7 hours of observations passing the data quality selection. 
This data set was analyzed using Model Analysis \cite{denaurois} with standard cuts. 
A significant excess of 225 events (7.8~$\sigma$) from the direction of this source was found.

\begin{figure}[!]
  \centering \includegraphics[width=8cm]{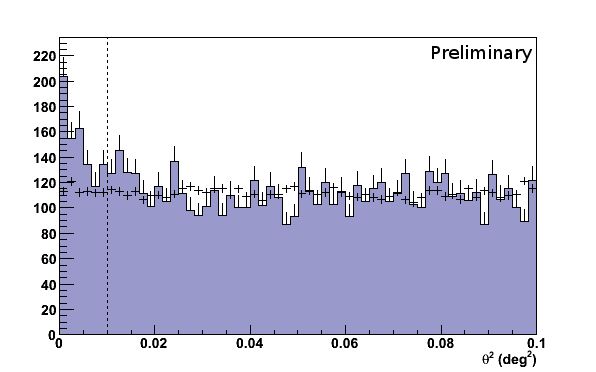}
  \centering \includegraphics[width=7cm]{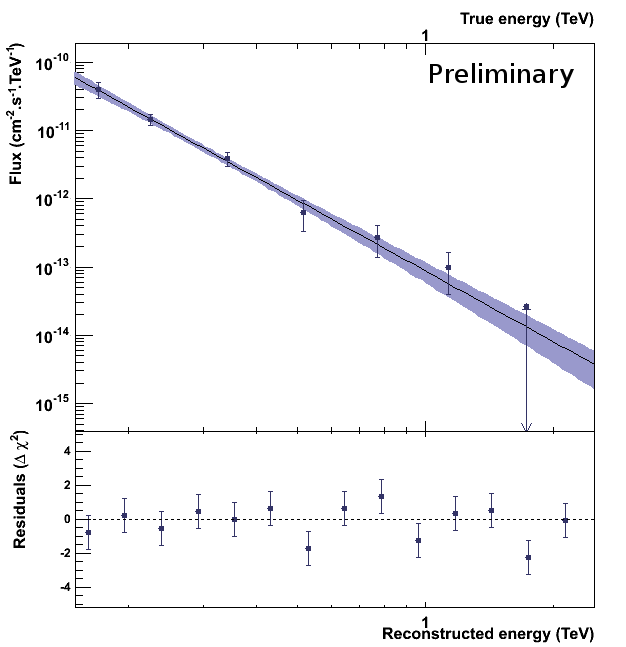}
  \caption{Left: Distribution of the $\theta^2$ for the on-source events (shaded) 
    and the normalized off-source events (crosses). 
    The dashed vertical line indicates the cut on the $\theta^2$ applied.  
 Right: VHE differential energy spectra of 1ES~0414+009. 
 The shaded band corresponds to the range of the power-law fit 
(68 $\%$ confidence level), taking into account statistical errors. 
Points are derived from the residuals in each energy bin reported in the bottom panel.}
  \label{fg1}
\end{figure}

The fit of the excess map with a 2D-function indicates that the source is 
point-like and is located at $\alpha_{J2000}$=$4h16m52.96s\pm15.3^{s}_{stat}$,$\delta_{J2000}=1^o5'20.4^"\pm15.1^{s}_{stat}$, 
consistent with the nominal position $\alpha_{J2000}$=$4^h16^m52.8^s$, $\delta_{J2000}=1^o5'24^"$  
\cite{ulmer83}. The  ON-source and normalised OFF source $\theta^2$ distribution (i.e. the distribution of the square of the angular 
difference between the reconstructed shower position and the source position) are shown in the Left panel of  
Fig.~\ref{fg1}.

The differential energy spectra (Right panel in Fig.~\ref{fg1}) was derived above 
the energy threshold of $\sim$150 GeV using excess events obtained 
with loose cuts analysis applied only to high-quality 4-telescope observations. 
The spectrum is compatible with a power law distribution with a photon index 
$\Gamma$ = 3.44 $\pm$0.27$_{stat}$ $\pm$ 0.2$_{sys}$ 
and with an integral flux of $1.83\pm0.21_{stat}\pm0.37_{syst}$ $\times10^{-12}\,cm^{-2}s^{-1}$
for E $>$ 200 GeV ($\sim$0.6\% of the Crab flux). 
No evidence of curvature in the spectrum or spectral variability was found.

The H.E.S.S. night-wise light curve shown in \ref{fg3} (upper panel) shows no 
evidence of variability ($\chi^2$$/$ndf = 78.7$/$79). 
No evidence of significant variability at any timescale has been found. 

\section{Multi-wavelength observations} \label{MWLDATA}

\begin{figure}[t!]
  \centering \includegraphics[width=12cm]{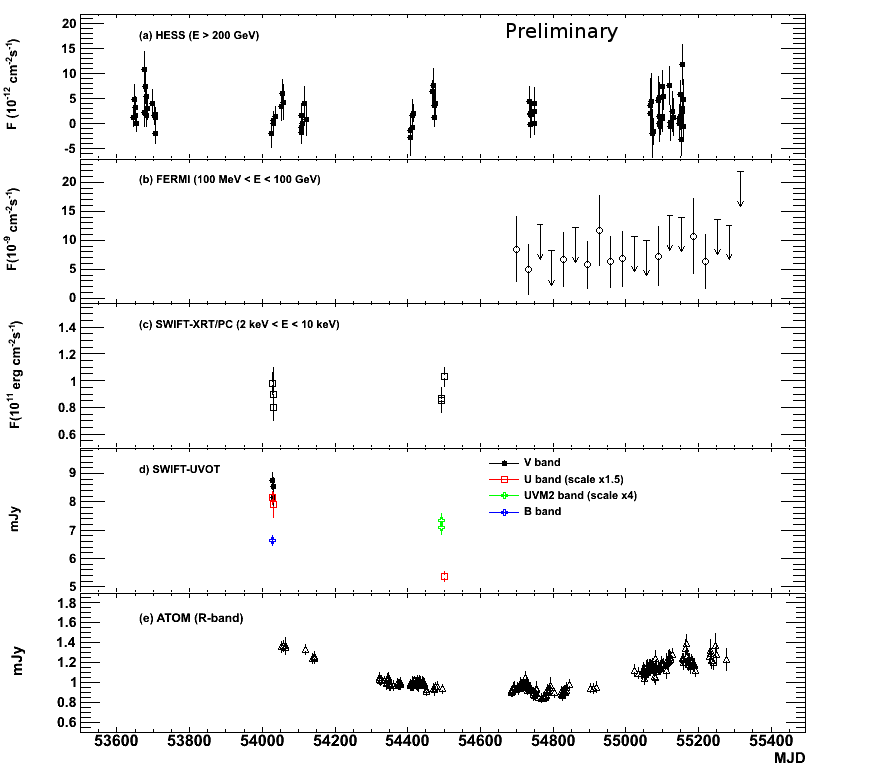}
  \caption{Light curve of the multi-wavelength observations of 1ES~0414+009 covering the period from 2005 to 2009.
Night-wise light curve of 1ES 0414+009 by H.E.S.S. (a),
{\textit SWIFT} XRT (c), UVOT (d), and ATOM (e) 
observations. The \textit{Fermi} light curve (b) has
32.5 day binning with 95\% upper limits.}
  \label{fg3}
\end{figure}

\begin{figure}[!]
  \centering \includegraphics[width=12cm]{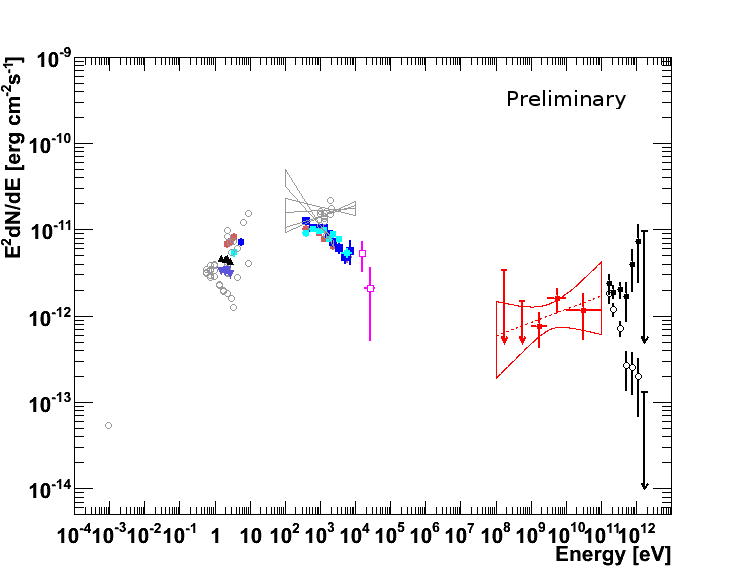}
  \caption{Average SED of 1ES~0414+009 with observations carried out between 2005 and 2009: 
HESS (black fill/open circles with/without EBL correction respectively); 
FERMI (red), SWIFT-BAT in 5 yrs (magenta), 
SWIFT XRT\&UVOT (purple October 2006, blue January 2008 and cyan February 2008) and ATOM (triangles). 
Gray points archival data from \cite{costa_ghisellini}.
}
  \label{fg4}
\end{figure}

\noindent {\bf HE Gamma-ray observations} The LAT telescope 
\cite{REF::LAT_INSTRUMENT} on the \textit{Fermi} Gamma-ray Space Telescope
detected the source at TS=68 (~7.9 $\sigma$) 
in 651 days (04 August 2008-17 May 2010). 
The energy spectrum is well described by a 
power law with spectral index of 1.76 $\pm$ 0.18 (see Fig.~\ref{fg4}). 
The integrated flux between 100 MeV and 100 GeV is 
(4.3 $\pm$ 2.2) $\times$ 10$^{-9}$ cm$^{-2}$ s$^{-1}$. 
No evidence of variability was found in the 32.5 day bins light curve 
(2nd panel in Fig.~\ref{fg3}).

\noindent {\bf X-ray observations} X-ray {\textit Swift}-XRT \cite{xrt} performed 3 observations 
in October 2006 and 3 between January and February 2008 (3rd panel \ref{fg3}). 
The spectra between 0.3-10 keV have been determined assuming a power-law model with the galactic absorption 
fixed at 9.15 $\times$ 10$^20$ cm$^{-2}$ \cite{elvis99}. The spectral indexes for 
the threes observations result to be around 2.2$-$2.3. 
Modest spectral variations among the different epochs have been observed: 
on Feb. 2008 the flux increased by $\sim$15\%, with a slight hardening of the spectrum ($\Delta \Gamma\approx$ 0.15).
The source was only marginally detected (2.5~$\sigma$) in the 14-30 keV band by 
the \textit{Swift} Burst Alert Telescope (BAT) \cite{bat} from 2004 to 2010.
The corresponding fluxes are reported in Fig.~\ref{fg4}. 

\noindent {\bf UV observations}: The UVOT instrument \cite{Roming} on board Swift 
observed simultaneously to the X-ray telescope the following UV and optical bands: 
V, U, UVM2, B (4th panel in Fig.~\ref{fg3}). As in the X-ray band, no significant variability was found in the UV band
in October 2006, nor in January 2008  (see fourth panel in Fig.~\ref{fg3},
where fluxes are reported with no extinction correction).

\noindent {\bf Optical observations}: The ATOM telescope \cite{atom}, located at the HESS site in Namibia, 
monitored the flux in the 4 different filters R, B, I and V. 
A slight decrease of the flux in 2008 was observed in the R-band (5th panel Fig.~\ref{fg3}).

\section{EBL constraints}\label{EBL}

With a redshift of 0.287 the reconstructed VHE spectrum of 1ES~0414+009 is expected to 
be strongly sensitive to the EBL density. 
Neverthless, 1ES~0414+009 does not provide stricter 
upper limits on the EBL level. It confirms the limits 
set by other HBL  \cite{nature,0347,1101} and indicates a low EBL level around 1-3 mm \cite{0347} 
close to the lower limits from galaxy counts.

Assuming as EBL model one of the best current estimate by \cite{franceschini}, 
both the \textit{Fermi} and H.E.S.S. absorption-corrected spectra are harder than $\Gamma=2$,  
indicating a high-energy peak above few TeV. 
The fit with a single power law of the FERMI-H.E.S.S. combined spectra has a spectral 
index of 1.84 $\pm$ 0.06. Therefore, 1ES~0414+009 can be classified as a hard TeV spectra objects.

\section{Broad-band SED}\label{SED}

The peak in the synchrotron emission is around 0.1 keV, while the peak in the high energy emission 
is above few TeV. This finding is difficult to explain with a standard one-zone SSC scenario, 
because Klein-Nishina effects tend to steepen the VHE spectrum unless the electron cooling is 
dominated by Compton scattering in the deep Klein-Nishina regime \cite{moderski2005}. 
The lack of hardening of the synchrotron spectrum at high X-ray energies (see BAT measurements) 
seems to disfavor the latter case. However, imposing 
the Thomson conditions for the two peaks requires either extreme parameters 
($\delta$ $>$200 and B$<$0.002 G) or a large emitting region 
(R$\sim$10$^{17}$ cm vs typical size around R$\sim$10$^{16}$ cm).

\section{Conclusion}\label{Conclusion}

The H.E.S.S. array of Cherenkov telescopes has detected a significant 
VHE $\gamma$-ray emission in the direction of 1ES~0414+009.
With its average flux of $\sim$0.6\% Crab Nebula flux above E $>$ 200 GeV, 1ES~0414+009 
results to be one of the most faint 
extra-galactic sources detected in the TeV domain. 
This source is also detected with the \textit{Fermi}-LAT instrument in the first 20 months of its operation between
2008 and 2010 and is also very faint also in the HE domain.

The VHE energy spectrum  is consistent with the current limit of the EBL and confirms the
low level of EBL around few $\mu$m given by the galaxy counts. 
The HE and VHE spectra (absorption corrected with an EBL model close to the lower limits) 
show a best-fit power law with an index harder than 2, indicating that 1ES~0414+009 can  
be classified among the hard-TeV BL~Lac objects. 

The overall SED of this source is averaged over five years and presents an 
IC peak energy above 1-2 TeV, which 
is difficult to be accounted for in the framework of a pure one-zone SSC model, 
unless using uncomfortable values for the main parameters. With the ten-fold sensitivity,
the future Cherenkov observatory CTA (Cherenkov Telescope Array) will 
provide more accurate measurements of the energy spectrum of distant blazars, such as 1ES~0414+009 
thus constraining the IC peak and helping to address the physics issues 
posed by the hard-TeV BL~Lac objects.

\section*{Acknowledgements}

The support of the Namibian authorities and of the University of Namibia
in facilitating the construction and operation of H.E.S.S. is gratefully
acknowledged, as is the support by the German Ministry for Education and
Research (BMBF), the Max Planck Society, the French Ministry for Research,
the CNRS-IN2P3 and the Astroparticle Interdisciplinary Programme of the
CNRS, the U.K. Science and Technology Facilities Council (STFC),
the IPNP of the Charles University, the Polish Ministry of Science and 
Higher Education, the South African Department of
Science and Technology and National Research Foundation, and by the
University of Namibia. We appreciate the excellent work of the technical
support staff in Berlin, Durham, Hamburg, Heidelberg, Palaiseau, Paris,
Saclay, and in Namibia in the construction and operation of the
equipment.\\

The $Fermi$ LAT Collaboration acknowledges support from a number of
agencies and institutes for both development and the operation of the
LAT as well as scientific data analysis. These include NASA and DOE in
the United States, CEA/Irfu and IN2P3/CNRS in France, ASI and INFN in
Italy, MEXT, KEK, and JAXA in Japan, and the K.~A.~Wallenberg
Foundation, the Swedish Research Council and the National Space Board
in Sweden. Additional support from INAF in Italy and CNES in France
for science analysis during the operations phase is also gratefully
acknowledged.

\end{document}